\documentclass[prl,showpacs,superscriptaddress,twocolumn,floatfix,amsmath,amssymb]{revtex4}
\usepackage[latin1]{inputenc}
\usepackage[T1]{fontenc}
\usepackage[english]{babel}
\usepackage{graphicx}
\usepackage{psfrag}
\usepackage{amsmath}
\usepackage{amssymb}
\usepackage{amscd}
\usepackage{eucal}
\usepackage{color}
\usepackage{bm}

\begin{document}
\title{Geometric Correlations and Breakdown of Mesoscopic Universality in
Spin Transport}
\author{ \.{I}. Adagideli}
\affiliation{Faculty of Engineering and Natural Sciences, Sabanci University, Orhanli-Tuzla, 34956 Istanbul, Turkey}
\author{Ph. Jacquod}
\affiliation{Physics Department,
   University of Arizona, 1118 E. 4$^{\rm th}$ Street, Tucson, AZ 85721, USA}
\author{M. Scheid}
\affiliation{Institut f\"ur Theoretische Physik, Universit\"at Regensburg, D-93040 Regensburg, Germany}
\author{M. Duckheim}
\affiliation{Dahlem Center for Complex Quantum Systems, Freie Universität
  Berlin,
  14195 Berlin, Germany}
\author{D. Loss}
\affiliation{Department of Physics, University of Basel, CH-4056
Basel, Switzerland}
\author{K. Richter}
\affiliation{Institut f\"ur Theoretische Physik, Universit\"at Regensburg, D-93040 Regensburg, Germany}
\date{\today}
\begin{abstract}
We construct a unified semiclassical theory of charge and spin
transport in chaotic ballistic and disordered diffusive mesoscopic
systems with spin-orbit interaction. Neglecting dynamic effects of
spin-orbit interaction, we reproduce the random matrix theory
results that the spin conductance fluctuates universally around zero
average. Incorporating these effects in the theory, we show that
geometric correlations generate finite average spin conductances,
but that they do not affect the charge conductance
to leading order. The theory,
which is confirmed by numerical transport calculations,
allows us to investigate the entire range from the weak to
the previously unexplored strong spin-orbit regime, where
the spin rotation time is shorter than the momentum relaxation time.
\end{abstract}

\maketitle{}
At low temperatures, linear electric transport
properties of complex mesoscopic systems are statistically
determined by the presence of few symmetries only, most notably
time-reversal and spin rotational symmetry~\cite{als,Bee97}. This
character of universality is believed to be independent of the
source of scattering in the system, and to exist in both ballistic
chaotic quantum dots or diffusive disordered
conductors~\cite{efetov}. Universality in electric transport holds
not only for global properties such as the conductance, but also for
correlators of transmission amplitudes between individual channels.
Thus, it is natural to expect that all transport properties
that depend solely on the scattering matrix are universal as well.
This conjecture has been theoretically verified for all charge
transport properties, under the sole assumption that scattering
generates complete ergodicity. Inspired by Ref.~\cite{Bar07},
several recent theoretical works~\cite{Naz07,Kri08,Ada09}  have
further suggested that spin transport in mesoscopic systems with
spin-orbit interaction (SOI) also displays universal random matrix theory (RMT)
behavior.
The agreement between numerics for a
disordered lattice~\cite{guo} and the RMT
prediction~\cite{Bar07} for the mesoscopic fluctuations of the spin
Hall conductance indeed seems to imply that RMT universality also exists in
magnetoelectric transport.

\begin{figure}
\includegraphics[width=\columnwidth]{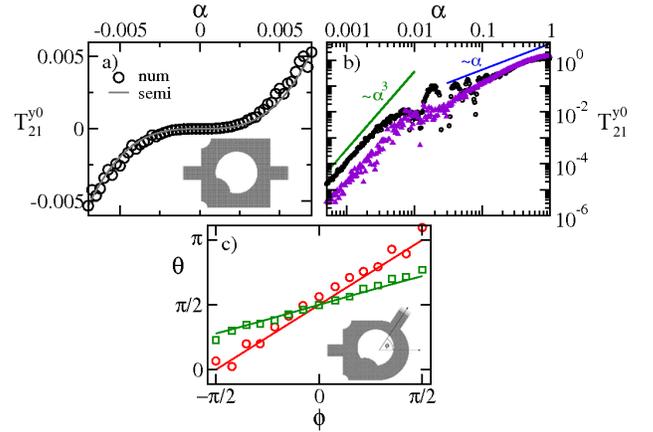}
\caption{\label{fig1} (Color online) Spin-dependent transmission
coefficients $\mathcal{T}_{21}^{y0}$,
Eq.~(\ref{tr_prob_semicl}), for (a) weak and (b) extended
range of SO coupling showing the crossover from cubic (green line) to linear (blue line)
behavior for the two-terminal chaotic quantum dot shown in the
inset of panel (a). (c) Spin current polarization angle
$\theta =\arctan ({\mathcal T}^{y0}_{21} / {\mathcal T}^{x0}_{21})$,
for the system shown in the inset, where the right lead encloses
an angle of $\phi$ with the $x$-axis in the linear (cubes) and the cubic
(circles) regimes.}
\end{figure}

In this work, we go beyond the conventional semiclassical theory of transport and show
that even when all requirements for universality are met and the fluctuations of the
spin and charge conductance as well as average charge conductance remain universal,
the average spin conductance is finite in disagreement with the RMT prediction.
This effect originates from the spin-orbit coupling through which the
electron spin perturbs the electron dynamics in such a fashion that, certain dynamical
correlations survive despite the self-averaging nature of ergodic dynamics.
These correlations depend on the geometry of the system, namely the relative positions
of the leads connecting the system to external
electronic reservoirs as well as the form of the SOI. As an example,
we consider a two-dimensional quantum dot with
Rashba SOI~\cite{Ras60} and find that the average two-terminal
spin-conductance $G^\mu$ is proportional to $(\hat{\bf z} \times {\bf R})_\mu$.
Here the vector ${\bf R}$ connects the two terminals,  $\hat{\bf z}$ is the unit
vector perpendicular to the dot and $\mu$ is the spin
component. This is illustrated in Fig.~\ref{fig1}(a) for the
corresponding spin transmission.
The polarization of the average spin current
is thus determined by the direction of the average electronic flow.
In bulk diffusive systems, when the mean free path is shorter than the spin rotation
length, this effect reduces to the extraction of the current-induced spin
accumulation (CISA) and the spin Hall effect~\cite{she_edge,Sih06,SpinBC} in finite systems.
We stress however that the consequences of these geometric correlations have
been considered in neither charge nor spin-transport in quantum dots.
Moreover, our calculations extend the existing theory for CISA and SHE in finite diffusive
systems to the strong SOI regime (i.e.~mean free path is longer than the spin rotation
length).
It is of practical importance to point out that the process that
leads to finite spin conductance is robust against temperature smearing or
dephasing. From the point of view of mesoscopic spintronics, this opens up possibilities
towards an electrically controlled generation and detection of pure spin currents,
since the uncontrolled mesoscopic fluctuations are suppressed by
simply raising the temperature.

We consider a mesoscopic quantum dot with no particular spatial symmetry
as sketched in the insets of
Fig.~\ref{fig1}. We treat impurity and boundary scattering on equal footing and consider diffusive as well as ballistic
chaotic charge dynamics.
The dot is connected to two or more
external leads.  For simplicity, we assume idealized reflectionless leads in which the SOI vanishes. The realistic
case of finite SOI in the leads can then be obtained by combining the scattering matrices of the realistic leads with
that of the quantum dot. This choice allows us to uniquely define transport spin currents through a cross-section
of the leads without the ambiguities that plagued bulk calculations~\cite{Ras03}.
The leads are maintained at different electrochemical potentials $eV_i$,
but have no spin accumulation. The scattering approach to transport gives
the spin and charge currents in lead $i$ as~\cite{But86}
\begin{equation}\label{scatt0}
I_i^{\mu} = \frac{e^2}{h}  \sum_{j} \mathcal{T}^{\mu0}_{ij} (V_i-V_j),
\end{equation}
with the generalized, spin-dependent transmission coefficients $\mathcal{T}_{ij}^{\mu\nu}$ obtained by
summing over all transport channels in leads $i$ and $j$~\cite{Bar07,Ada09},
\begin{equation}\label{eq:tr_prob}
\mathcal{T}_{ij}^{\mu\nu} =\sum_{m \in i,n \in j}{\rm tr} [ t_{mn}^\dag
\sigma_{\mu} t_{mn}\sigma_\nu], \;\;\;\; \mu,\nu = 0,x,y,z \,.
\end{equation}
Here, $\sigma_\mu$ are Pauli matrices ($\sigma_0$ is the identity matrix)
and the trace is taken over the spin degree of freedom.
The transmission amplitudes
in Eq.~(\ref{eq:tr_prob}) can be expressed in terms of the Green's function~\cite{Bar_Stone}.
Next, we obtain
the full Green's function $G^R({\bf r},{\bf r}')$ by
either (i)~the conventional Born approximation for impurity
scattering inside the conductor or (ii)~by a multiple reflection expansion
for boundary scattering~\cite{BB70,Ada10}. In case (ii),
$G^R({\bf r},{\bf r}')$ is expressed as an iterative solution of
\begin{equation}
G^R({\bf r},{\bf r}')=G^R_0({\bf r},{\bf r}')-
2\int \! d\alpha \,\partial
G^R_0({\bf r},{\bf \alpha})
G^R({\bf \alpha},{\bf r}'),
\label{EQ:MRE}
\end{equation}
where
$\partial G^R_0({\bf r},\alpha)={\bf \hat{n}}_\alpha\cdot \nabla G^R_0({\bf r},{\bf x})\vert_{{\bf x}={\bf \alpha}}$,
with ${\bf \hat{n}}_\alpha$ the (inner) unit normal vector at the boundary point $\alpha$.
Finally, we evaluate the surface integrals in Eq.~(\ref{EQ:MRE})
asymptotically
as
$k_F L\rightarrow \infty$, where $k_F$ is the Fermi
wavenumber and $L$ is the linear size of the conductor~\cite{Ada10}.
We obtain
\begin{eqnarray}\label{tr_prob_semicl}
{\mathcal T}_{ij}^{\mu0} &=&  \!\! \int_i \! {\rm d} y \!\int_j \! {\rm d} y_0
\sum_{\gamma,\gamma'}
A_\gamma A_{\gamma'}^* e^{i(S_\gamma-S_{\gamma'})}
{\rm tr}[V_\gamma \sigma_\mu V_{\gamma'}^\dagger],
\end{eqnarray}
where the sums run over all trajectories $\gamma$ starting at $y_0$ on a cross-section of
the injection lead and ending at $y$ on the exit lead.
The classical action of $\gamma$ is $S_{\gamma}$ in units of $\hbar$ and
its stability is given by $A_\gamma$ which includes a prefactor $(2 \pi i \hbar)^{-1/2}$ as well as
Maslov indices. For the spin
dependent part, we specialize to the Rashba SOI
$ H_{\rm R} = (\hbar k_\alpha / m) (p_x \sigma_y - p_y \sigma_x) $,
where $k_\alpha^{-1}$ is the spin precession length~\cite{Ras60}.
We then obtain
\begin{eqnarray}\label{eq:vmatr}
V_\gamma & = & \prod_{i=1}^{N_\gamma} V_{i,\gamma}=
\prod_{i=1}^{N_\gamma}
U_{i,\gamma}(1 + \delta U_{i,\gamma} + \xi \, \delta U_{i,\gamma}^{\rm hw}) \\
\delta U_{i,\gamma} & = &
\frac{ k_\alpha }{4 k_{\rm F}} \left(\frac{\sin(k_\alpha|{\bf r}_i|)}{k_\alpha|{\bf r}_i|}-1
\right) \boldsymbol{\eta} \cdot {\bf \hat{r}}_i
\,
\label{eq:umatr}
\\
\delta U_{i,\gamma}^{\rm hw} & = & \frac{ k_\alpha }{2 k_{\rm F}}
\left(\frac{\sin(k_\alpha|{\bf r}_i|)}{k_\alpha|{\bf r}_i|}-1
\right)
\left(\boldsymbol{\eta} \cdot {\bf \hat{r}}_i -
\frac{\boldsymbol{\eta} \cdot {\bf \hat{n}}_i}{\cos\theta_i}
\right) \nonumber \\
&&\quad + \frac{\sigma_z \sin\theta_i}{2 k_{\rm F}|{\bf r}_i|\cos\theta_i}(1-\cos(k_\alpha|{\bf r}_i|)) \, .
\label{eq:umatr-hw}
\end{eqnarray}
Here $\xi=0$ for a disordered system with weak, short-ranged impurities and
$\xi=1$ for a ballistic quantum dot with hard-wall confinement or a disordered system with
strong,
extended impurities. In both cases $\gamma$ consists of segments ${\bf r}_i=(x_i,y_i,0)$
with
$i=1,2,...N_\gamma$ , $\hat{\bf r}_i={\bf r}_i/|{\bf r}_i|$,
${\bf \hat{n}_i}$ is the (inner) unit normal vector and $\theta_i$ is the angle of incidence at
the $i$th reflection point,
$\boldsymbol{\eta}=\hat{\bf z} \times \boldsymbol{\sigma}$ and
$U_{i,\gamma}=\exp[-i k_\alpha \boldsymbol{\eta}
\cdot {\bf r}_i/2]$ is the Rashba spin rotation
matrix along that segment.
We note that there are also corrections to $A_\gamma$
which we have already ignored here, because they do not contribute
to the spin conductance (however they generate diffractive corrections
to the charge conductance).
The Eqs.~(\ref{tr_prob_semicl}-\ref{eq:umatr-hw}) completely describe spin and
charge dynamics of coherent
conductors.

The conventional semiclassical theory is obtained via the approximation
$V_\gamma \approx \prod_{i=1}^{N_\gamma} U_{i,\gamma}$, which leads to the
universal RMT predictions for charge transport~\cite{Mat92,Zaitsev05}. We now show that
this approximation also leads to RMT results for spin transport for $\mu\ne 0$. We
first start from
the diagonal approximation, where $\gamma=\gamma'$, and obtain
${\rm tr}[V_\gamma \sigma_\mu V_{\gamma}^\dagger]=0$,
showing that the diagonal contribution to
the spin current vanishes.
The next-order contributions within the conventional semiclassical theory of transport are the
loop corrections, in which a self-crossing trajectory $\gamma$, is paired with a
path $\gamma'$ avoiding the crossing and going around the loop in the
the opposite direction~\cite{Ale96,Ric02}. Along the loop, $\gamma'$ is the time-reversed
of $\gamma$, and
the loop contributions are proportional to
$\langle \mathrm{tr} [ U_{\gamma_l} \sigma_\mu U_{\gamma_l} ] \rangle$ ,
where $U_{\gamma_l}$ gives the spin rotation along the loop only.
For large SOI, $U_{\gamma_l}$ is random, thus averaging produces
vanishing weak localization correction to the spin conductance.
For weaker SOI, we expand all
spin rotation angles to second order in $k_\alpha L$ to obtain
$\big\langle\mathrm{tr} [ U_{\gamma_l} \sigma_\mu U_{\gamma_l} ]
\big\rangle \approx 2 i \delta_{\mu
z}\big\langle\sin\big(k_\alpha^2 \, \delta\mathcal{A}_{\gamma_l}\big)\big\rangle \, .
$
The area difference $\delta \mathcal{A}_{\gamma_l}$ is given approximately by
twice the directed area of the weak
localization loop. For a
chaotic system, the areas are symmetrically distributed around zero,
thus the average vanishes.
We note that extending the semiclassical approach
of Ref.~\cite{Bro06} to the calculation of the variance of the spin conductance,
one straightforwardly reproduces the leading-order RMT results of Ref.~\cite{Bar07}.
Details of this calculation will be presented elsewhere~\cite{Ada10}.
We conclude that conventional semiclassical theory,
which neglects effects of spin on the charge dynamics, only reproduces RMT predictions.

We next include the effects of SOI on the electronic dynamics and consider a
two-dimensional
conductor which can be either
a ballistic quantum dot with hard-wall confinement, or
a disordered system with short-ranged
impurities.
To do this, we go back to Eqs.~(\ref{eq:vmatr}-\ref{eq:umatr-hw}) and
include the corrections to the amplitude
$A$ and the spin matrix $U$ to order
${\cal O}(k_\alpha/k_{\rm F})$ and ${\cal O}(1/k_{\rm F} |{\bf r}_i|)$ .
We now assume that different trajectory segments are uncorrelated
and define ${\cal U}_{l,\gamma}  =
\prod_{i=l+1}^{N_\gamma} U_{i,\gamma}$ to obtain
\begin{eqnarray}
\label{spin_cond_avg}
\langle {\rm tr} [V_{\gamma} \, V_{\gamma}^\dagger \, \sigma_\mu ]\rangle
= \left\langle \sum_{l=1}^{N_\gamma}
{\rm tr} \left[{\cal U}_{l,\gamma} \, V_{l,\gamma} V_{l,\gamma}^\dagger \,
{\cal U}_{l,\gamma}^\dagger \, \sigma_\mu \right]  \right\rangle_\gamma \, .
\label{EQ:spin_gen}
\end{eqnarray}
We see that spin currents have contributions from every trajectory segment,
which are further rotated by the fluctuating spin-orbit
fields of the subsequent reflections. We
distinguish three different regimes that depend on the balance between
linear system size
$L$, the mean distance between (boundary or impurity) scatterings
$\ell=\langle |{\bf r}_l|\rangle$, and SOI length $k_\alpha^{-1}$:
(i)~the spin-ballistic small
SOI limit $k_\alpha L, k_\alpha \ell \ll 1$, (ii)~the spin-diffusive limit
$ k_\alpha \ell \ll 1 \ll k_\alpha L$, (iii)~ the spin-chaotic strong SOI limit
$k_\alpha L, k_\alpha \ell \gg 1$. In regimes (i) and (iii),
the orbital dynamics can be chaotic ballistic or
diffusive depending on the ratio between $\ell$ and $L$.
We will be focusing on long ergodic or diffusive trajectories $\gamma$ for which we
ignore the averages  $\langle \sin\theta_i \rangle_\gamma$ and
$ \langle{\bf \hat{n}}_i\rangle_\gamma$ for all three regimes, save for the case of a
quantum dot in regime (i) (see below).

In the {\em small SOI regime (i)}, we expand the rhs of
Eq.~(\ref{spin_cond_avg}) to leading
order in $k_\alpha \ell$ setting ${\cal U}_{l,\gamma} = \openone$
and $1-\sin (k_\alpha |{\bf r}_l|)/k_\alpha |{\bf r}_l|
\simeq (k_\alpha |{\bf r}_l|)^2/6$ in
Eqs.~(\ref{eq:umatr},\ref{eq:umatr-hw}). We get
\begin{equation}
\langle {\rm tr} [V_{\gamma} \, V_{\gamma}^\dagger \, \sigma_\mu ]\rangle
\approx \frac{k_\alpha^3(1+2\xi)}{6 k_{\rm F}}\left\langle
\sum_{l=1}^{N_\gamma} |{\bf r}_l|
(\hat{\bf z}\times{\bf r}_l)_\mu \right\rangle_\gamma \, .
\end{equation}
We now perform the average $\langle \ldots \rangle_\gamma$ over
the set of trajectories $\gamma$.
Although individual ${\bf r}_i$ are
pseudorandom in length and direction,
being generated by the cavity's chaotic dynamics,
they satisfy  $\sum_i {\bf r}_i^\gamma\approx {\bf R}_{ij}$,
where ${\bf R}_{ij}$ is the $\gamma$-independent vector connecting the
injection
and exit terminal. We thus obtain
$ \langle {\rm tr} [V_{\gamma} \, V_{\gamma}^\dagger \, \sigma_\mu ]\rangle
= C [k_\alpha^3 \ell(1+2\xi)/(3 k_{\rm F})]
(\hat{\bf z}\times{\bf R}_{ij})_\mu \, , $
where $C$ is a number of order one that depends on geometric
details of the cavity. This factor multiplies the independently averaged
orbital terms in Eq.~(\ref{tr_prob_semicl}) for $\gamma=\gamma'$, which we compute
as in, e.g. Ref.~\cite{Ric02}. We estimate $\ell=\langle |{\bf r}_l|\rangle \simeq \pi {\cal A/L}$
for a chaotic dot of area ${\cal A}$ and perimeter ${\cal L}$, and
$\ell= v_F \tau$ for a diffusive system with momentum relaxation time $\tau$.
We finally obtain
\begin{eqnarray}\label{eq:spin_ballistic}
\langle {\mathcal T}^{\mu0}_{ij} \rangle\! &=& \! C \,
\frac{k_\alpha^3 \ell (1+2\xi)}{3 k_{\rm F}}
 (\hat{\bf z}\times{\bf R}_{ij})_\mu \! \times \!\left\{
\begin{array}{cc}
\frac{N_i N_j}{N_{\rm T}} & , \ell \gtrsim L \, , \\
\frac{k_{\rm F} W \ell}{ L} & , \ell \ll L \, ,
\end{array}
\right. \qquad
\end{eqnarray}
with the number $N_i = {\rm Int}(k_{\rm F} W_i/\pi)$ of channels in
lead $i$,  $N_{\rm T}=\sum_i N_i$ and $W={\rm min}W_i$ the width of the narrowest
lead. In the ballistic limit, this formula has a correction term
$\frac{k_\alpha^3 \ell^2\xi N_i N_j}{3 k_{\rm F}N_{\rm T}^2}\sum_l N_l (\hat{\bf z}\times\hat{\bf R}_l)_\mu$,
where ${\bf R}_l$ is the average momentum direction of electrons entering through lead $l$,
originating from nonzero $ \langle{\bf \hat{n}}_i\rangle_\gamma$~\cite{Ada10}.
We see that the average
spin-dependent transmission, and thus the average spin currents, are
determined by the relative position of the injection and exit lead
and are proportional to the classical conductance from $j$ to $i$.

In the {\em spin-diffusive case (ii)}, $L \gg k_\alpha^{-1} \gg
\ell$,
the spins precess around randomly oriented SOI fields, thus
relaxing via the Dyakonov-Perel mechanism. In particular, we can no longer set
${\cal U}_{l,\gamma}= \openone$  in Eq.~(\ref{EQ:spin_gen}).
Instead, we assume that $\gamma$ is a stochastic sequence of segments
with random orientations $\varphi_i$, which determine the spin rotation
$U_{i,\gamma}$. The sequence of rotations is computed by averaging
over $\varphi_i$. For a general Pauli spin matrix
${\bf s} \cdot \boldsymbol{\sigma}$ one has
\begin{eqnarray}
\label{angle_avg}
\int \frac{d\varphi_i}{2 \pi} U_{i,\gamma} \, {\bf s} \cdot \boldsymbol{\sigma}
U_{i,\gamma}^\dag &=&\cos^2(k_\alpha |{\bf r}_i|/2) \, {\bf s}\cdot \boldsymbol{\sigma}  \\
&&\! +\! (|{\bf r}_i|^2/2)  \sin^2(k_\alpha |{\bf r}_i|/2) \, \boldsymbol{\eta} \,
({\bf s}\cdot \boldsymbol{\sigma}) \,
\boldsymbol{\eta} \,.\nonumber
\end{eqnarray}
This average is different for in-plane
and out-of-plane polarization, which is the origin of the
anisotropy of the Dyakonov-Perel spin-relaxation time.
In our case, the generated spin is in-plane and the second term in Eq.~(\ref{angle_avg})
vanishes~\cite{she_caveat}.
We have
\begin{equation}
\label{eq:17}
\frac{\langle V_{\gamma}\,V_{\gamma}^\dagger\rangle-1}{1+2\xi}
\approx -\left\langle \sum_{l=1}^{N_\gamma} e^{-k_\alpha^2 \ell
v_{\rm F} \tau_l} \frac{k_\alpha}{2 k_{\rm F}} \frac{k_\alpha^2 \ell}{6}
\boldsymbol{\eta}\cdot {\bf r}_l
\right\rangle_\gamma \, ,
\end{equation}
where we used $k_\alpha \ell \ll 1 \ll k_\alpha L$,
approximated $|{\bf r}_i| \approx \ell$, $\forall i$ and introduced the
duration $\tau_l$ of the first $l$ segments of $\gamma$.
For each possible choice of
$l$, the spin rotation thus separates into
a spin independent piece for segments $1,\cdots,l-1$,
a spin generation piece on segment $l$, and a spin relaxation piece
on segments $l+1,\cdots,N_\gamma$. Fixing
the endpoint ${\bf r}_l$ of segment $l$ and summing over all possible orbits
we obtain that the spin conductance is proportional to a product of
(i)~a diffusive probability $P({\bf x}_l,{\bf x}_j)$
to go from the injection lead
to ${\bf r}_l$, (ii)~a spin
generation factor
$(1+2\xi)k_\alpha^3 \ell \boldsymbol{\eta} \cdot ({\bf x}_l-{\bf x}_{l'})/12 k_{\rm F}$
multiplying the probability of
ballistic propagation from ${\bf x}_l$ to ${\bf x}_{l'}$,
(iii)~a diffusive probability to propagate from point ${\bf x}_{l'}$ to
the exit lead times the probability that the spin survives this diffusion.
Thus we have
\begin{eqnarray}
\langle \mathcal{T}^{\mu0}_{ij} \rangle &\propto& \epsilon_{3\mu\nu}
\frac{k_\alpha^3 \ell}{k_{\rm F}}
\int d{\bf x}_i \, d{\bf x}_j \, d{\bf x}_l \,
d{\bf x}_{l'}
P({\bf x}_l,{\bf x}_j) ({\bf x}_l-{\bf x}_{l'})^\nu \nonumber \\
& \times &  (1+2\xi) \frac{e^{-\vert {\bf x}_l-{\bf
x}_{l'}\vert/\ell}}{2\pi \vert {\bf x}_l-{\bf x}_{l'}\vert} P({\bf
x}_i,{\bf x}_{l'})e^{-k_\alpha \vert {\bf x}_{l'}-{\bf x}_i\vert}.
\end{eqnarray}
Since the length scale characterizing
$P({\bf x}_l,{\bf x}_j)$ is $L$, we evaluate the integrals above asymptotically in the
limit $k_\alpha \ell \ll 1 \ll k_\alpha L$. After some algebra we
finally obtain
\begin{eqnarray}\label{eq:regii}
\langle \mathcal{T}^{\mu0}_{ij} \rangle &\propto&
{\rm sgn}(k_\alpha) (1+2\xi)\frac{k_\alpha^2 \ell^2 W}{L^2} \,
(\hat{\bf z} \times {\bf R}_{ij}) \, ,
\end{eqnarray}
up to a factor of order unity depending on details of how the leads (with
width $W$) are attached to the cavity.
Noting that for our geometry ${\bf R}_{ij}$ is in the direction of the current flow and its magnitude is
$L$, we obtain that the spin conductivity is $\sigma_s\propto e k_\alpha^2 \ell^2$ in agreement with the
spin diffusion equation calculations~\cite{she_edge,SpinBC}.

{\em Spin chaos regime (iii)}:
Similar to regime (ii), we average over uncorrelated direction
angles $\theta_i$ but do not Taylor-expand $\sin (k_\alpha |{\bf r}_l|)/
k_\alpha |{\bf r}_l|-1$.
We instead take the average over the segment lengths $|{\bf r}|_i$
as $\prod_{i=l+1}^{N_\gamma} \langle \cos^2(k_\alpha |{\bf r}_i|/2) \rangle
\approx 2^{N_\gamma-l}$ in a chaotic/stochastic system with $k_\alpha L \gg 1$.
Eq.~(\ref{eq:17}) is then replaced by
\begin{equation}
\frac{\langle V_{\gamma}\,V_{\gamma}^\dagger \rangle-1}{1+2\xi}=
\left\langle
\sum_{l=1}^{N_\gamma}
2^{l-N_\gamma}
\frac{k_\alpha}{2k_{\rm F}}
\left(
\frac{\sin (k_\alpha |{\bf r}_l|)}{k_\alpha |{\bf r}_l|}-1
\right)
\boldsymbol{\eta}\cdot \hat{\bf r}_l
\right\rangle_\gamma \, .
\nonumber
\end{equation}
Averaging over
$\gamma$ we see that the dominant contribution is the last term.
We thus approximate
the sum by its last term, and take $k_\alpha |{\bf r}_{N_\gamma}| \simeq
k_\alpha L \gg 1$ to obtain
$\langle V_{\gamma}\,V_{\gamma}^\dagger \rangle_\gamma\!=\!  1\!+\!
( C' k_\alpha/2 k_{\rm F})  \boldsymbol{\eta}\cdot \hat{\bf R}_j $.
Here
$C'$ is $(1+2\xi)$ times a constant of order unity
that depends on the details of the scattering near the lead. We finally obtain the
transmission coefficient
\begin{eqnarray}\label{eq:regiii}
 \langle \mathcal{T}^{\mu0}_{ij} \rangle &= &
C'\frac{k_\alpha}{2 k_{\rm F}} (\hat{\bf z} \times \hat{\bf R}_j)_\mu
\times \left\{
\begin{array}{cc}
N_i N_j/N_{\rm T} & \ell \gtrsim L \, , \\
k_{\rm F} W \ell/L & \ell \ll L \, .
\end{array}
\right. \qquad
\end{eqnarray}

Equations (\ref{eq:spin_ballistic}), (\ref{eq:regii})
and (\ref{eq:regiii}) are our  main results. They show how
a finite spin conductance emerges from classical geometric correlations
depending on the positions of the leads. These equations
can be straightforwardly extended to Dresselhaus SOI by substituting
$\hat{\bf z} \times {\bf Q} \rightarrow (Q_x,-Q_y,0)$ for
${\bf Q}={\bf R}_{ij}$ [Eqs.~(\ref{eq:spin_ballistic}) and (\ref{eq:regii})]
or ${\bf Q}={\bf R}_j$ [Eq.~(\ref{eq:regiii})].

To check these predictions we performed numerical recursive Green's function
quantum transport calculations for a tight-binding
Hamiltonian~\cite{Wimmer09} with Rashba SOI and evaluated the spin-resolved
transmission probabilities between two leads as defined in Eq.~(\ref{eq:tr_prob})
for both the chaotic and diffusive cases.
We computed the transmission for chaotic cavities, shown as insets in
Fig.~\ref{fig1}, averaged over 2000 different configurations of the Fermi energy
and the position and orientation of the central antidot.
Panel a) shows for the small $\alpha=a \, k_\alpha$ regime (i)
that the numerically obtained ${\mathcal T}^{y0}_{21}$ (dots) for the cavity
in the inset agrees very well with the predicted cubic behavior,
Eq.~(\ref{eq:spin_ballistic}), (solid line) for $C\!=\!1$.
In panel b) ${\mathcal T}^{y0}_{21}$ is depicted
for the same chaotic cavity (black circles) and for a
square cavity with Anderson disorder (violet triangles) for the entire range
from weak to strong SOI (regime (i) to (iii)) demonstrating the
crossover from cubic to linear behaviour according to Eqs.\
(\ref{eq:spin_ballistic}) and  (\ref{eq:regiii}).
In panel c) we further numerically confirm the predicted direction of the in-plane
spin polarization
$\theta =\arctan ({\mathcal T}^{y0}_{21} / {\mathcal T}^{x0}_{21})$
for regime (i) (dashed line, Eq.~(\ref{eq:spin_ballistic})) and
regime (iii) (solid line, Eq.~(\ref{eq:regiii}))
by rotating the right lead around the semicircle billiard
shown in the inset.

In conclusion, we have presented a semiclassical calculation of spin
transport in mesoscopic conductors which incorporates next-to-leading order
corrections to the semiclassical Green's function. We showed that in contrast
to RMT predictions, the average spin conductance does not
vanish, even if all the conventional conditions for universality
are met. Our method moreover allowed us to investigate
the strong SOI regime for finite diffusive systems for the first time,
Eq.~(\ref{eq:regiii}).

This work has been supported by the funds of the Erdal \.In\"on\"u chair and by TUBA under grant I.A/TUBA-GEBIP/2010-1 (IA),
by NSF under grant DMR-0706319 (PJ) and by DFG within SFB 689 (MS,KR).
IA and PJ thank the University of Regensburg, and PJ thanks the Basel Center for Quantum Computing
and Quantum Coherence for their hospitality.


\begin{thebibliography}{99}
\bibitem{she_caveat} We note that in more complicated geometries such as side contacts
there will be out-of-plane spin polarization~\cite{she_edge}.
\end{thebibliography}
\end{document}